\documentclass[10pt,twocolumn,letterpaper]{article}

\usepackage[iccv]{iccv}      

\usepackage{colortbl}

\usepackage{multicol}
\usepackage{multirow}

\definecolor{iccvblue}{rgb}{0.21,0.49,0.74}

\usepackage[pagebackref,breaklinks,colorlinks,allcolors=iccvblue]{hyperref}
\def\paperID{2655} 
\def\confName{ICCV}
\def\confYear{2025}

\title{4DGS-CC: A Contextual Coding Framework for 4D Gaussian Splatting \\ Data Compression}

\author{Zicong Chen{$^{1}$},~ Zhenghao Chen{$^{2}$},~ Wei Jiang{$^{3}$},~ Wei Wang{$^{3}$},~ Lei Liu{$^{4}$},~ Dong Xu$^{4}$ \\
$^{1}$~Beihang University,~~{$^{2}$}~The University of Newcastle, Australia, ~~\\{$^{3}$}~Futurewei Technologies Inc,{$^{4}$}~~The University of Hong Kong~~ \\
\tt\small{chenzicong@buaa.edu.cn, zhenghao.chen@newcastle.edu.au, dongxu@hku.hk}}

\begin{document}
\maketitle



\newcommand{\MyMapTemplatePrefixc}[4]{\expandafter#1\csname#3#4\endcsname{#2{#4}}} 
\forcsvlist{\MyMapTemplatePrefixc {\def} {\mathcal}{c}} {A,B,C,D,E,F,G,H,I,J,K,L,M,N,O,P,Q,R,S,T,U,V,W,X,Y,Z}  

\newcommand{\MyMapTemplatePrefixtb}[5]{\expandafter#1\csname#4#5\endcsname{#2{#3{#5}}}} 
\forcsvlist{\MyMapTemplatePrefixtb {\def} {\tilde}{\mathbf}{t}} {A,B,C,D,E,F,G,H,I,J,K,L,M,N,O,P,Q,R,S,T,U,V,W,X,Y,Z}  
\forcsvlist{\MyMapTemplatePrefixtb {\def} {\tilde}{\mathbf}{t}} {0,1,a,b,c,d,e,f,g,h,i,j,k,l,m,n,o,p,q,r,s,u,v,w,x,y,z}  

\newcommand{\MyMapTemplateNoPrefix}[3]{\expandafter#1\csname#3\endcsname{#2{#3}}}
\forcsvlist{\MyMapTemplateNoPrefix {\def} {\mathbf} } {0,1,a,b,c,d,e, f, g, h, i, j, k, l, m, n, o, p, q, r, u, v, w, x, y, z} 
\forcsvlist{\MyMapTemplateNoPrefix {\def} {\mathbf} } {A,B,C,D,E,F,G,H,I,J,K,L,M,N,O,P,Q,R,S,T,U,V,W,X,Y,Z}  

\def\ba{\bm{\alpha}}
\def\bb{\bm{\beta}}
\def\bd{\bm{\delta}}
\def\bt{\bm{\theta}}
\def\bz{\bm{\zeta}}

\def\bP{\bm{\Phi}}
\def\bQ{\bm{\Psi}}
\def\bS{\bm{\Sigma}}

\def\bbR{{\mathbb R}}
\def\bbE{{\mathbb E}}

\def\tr{\mbox{tr}}
\def\Pr{{P}}

\def\etal{\emph{et al.}\@\xspace}
\def\ie{\emph{i.e.}\@\xspace}
\def\eg{\emph{e.g.}\@\xspace}
\def\resp{\emph{resp.}\@\xspace}

\definecolor{rowblue}{RGB}{220,230,240}

\begin{abstract}

%
\vspace{-3mm}

Storage is a significant challenge in reconstructing dynamic scenes with 4D Gaussian Splatting (4DGS) data. In this work, we introduce 4DGS-CC, a contextual coding framework that compresses 4DGS data to meet specific storage constraints.
Building upon the established deformable 3D Gaussian Splatting (3DGS) method, our approach decomposes 4DGS data into 4D neural voxels and a canonical 3DGS component, which are then compressed using Neural Voxel Contextual Coding (NVCC) and Vector Quantization Contextual Coding (VQCC), respectively.
Specifically, we first decompose the 4D neural voxels into distinct quantized features by separating the temporal and spatial dimensions. To losslessly compress each quantized feature, we leverage the previously compressed features from the temporal and spatial dimensions as priors and apply NVCC to generate the spatiotemporal context for contextual coding.
Next, we employ a codebook to store spherical harmonics information from canonical 3DGS as quantized vectors, which are then losslessly compressed by using VQCC with the auxiliary learned hyperpriors for contextual coding, thereby reducing redundancy within the codebook.
By integrating NVCC and VQCC, our contextual coding framework, 4DGS-CC, enables multi-rate 4DGS data compression tailored to specific storage requirements. Extensive experiments on three 4DGS data compression benchmarks demonstrate that our method achieves an average storage reduction of approximately \textbf{12$\times$} while maintaining rendering fidelity compared to our baseline 4DGS approach.
%

\end{abstract}

\vspace{-2mm}
\section{Introduction}
\vspace{-2mm}
Dynamic scene rendering is pivotal for applications such as virtual reality, augmented reality, gaming, and robotics, where real-time, precise depiction of moving objects and their environments is essential for an immersive experience. While advancements like Neural Radiance Fields (NeRFs)~\cite{mildenhall2021nerf} have enabled high-fidelity scene reconstruction, they often come with the drawback of slow rendering speeds. To overcome this limitation, 3D Gaussian Splatting (3DGS)~\cite{kerbl20233d} significantly accelerates rendering by representing static scenes with 3D Gaussian data, thereby achieving real-time performance.


Despite the tremendous success of 3DGS techniques in rendering static scenes, extending them to dynamic scenes using 4D Gaussian Splatting (4DGS) presents significant challenges, with high storage costs emerging as one of the primary concerns~\cite{luiten2024dynamic}. To achieve high fidelity, some approaches either directly model temporal transformations of 3D Gaussian data~\cite{sun20243dgstream, li2024spacetime, katsumata2024compact} or incorporate an additional time dimension~\cite{yang2023real, duan20244d} to generate 4D Gaussian data. However, these methods rely on extensive attributes and impose a substantial storage burden.

To produce a more compact 4DGS format, recent works typically leverage deformable 3DGS techniques to disentangle 4D Gaussian data into a canonical 3D Gaussian set and a Gaussian deformation field~\cite{yang2024deformable, huang2024sc, wu20244d, xu2024grid4d, yan20244d}. In this framework, the deformation field compactly encodes the Gaussian motion and deformation information for each time-step, ensuring that only one canonical 3DGS is maintained. For example, Wu~\etal~\cite{wu20244d} propose a method called 4DGS, which employs compact 4D neural voxels to represent the Gaussian deformation field using variables such as positions, scales and rotations.

While deformable 3DGS strategies employ compact structures to reduce 4DGS data storage, two major challenges persist. First, redundancy within the structure is not entirely eliminated. For example, even though the deformation field uses compact components such as 4D neural voxels~\cite{cao2023Hexplane}, unexploited spatiotemporal correlations within and across time-steps remain. Second, many practical applications, including mobile VR, require multi-rate compression that allows users to balance rendering quality and file size based on factors such as network bandwidth. This functionality is currently absent in existing deformable 3DGS methods, limiting their broader adoption.
To address these concerns, we propose a 4DGS data compression method to further exploits redundancy to push the boundaries of 4DGS compression. We follow the high-level syntax of deformable 3DGS, comprising a canonical 3D Gaussian set and 4D neural voxel components. To exploit redundancy within each component, we incorporate neural contextual coding techniques inspired by the success of neural image compression (NIC) methods~\cite{balle2018variational,balle2016end,minnen2018joint,6,1,12}.

For compressing the 4D neural voxels, which are constructed from multi-resolution grid structures (e.g., Hexplane~\cite{cao2023Hexplane}) with each grid containing a collection of features, we first apply quantization to achieve lossy compression. Next, we disentangle the quantized features into spatial and temporal dimensions. Specifically, the position of each plane $s$ in the coarse-to-fine structure defines the spatial axis, while each plane is partitioned into $t$ rows, with each row representing deformation information of the static scene at a specific time-step, inspired by the 4D neural voxel design~\cite{cao2023Hexplane}.
Finally, we introduce a Neural Voxel contextual Coding (NVCC) approach to losslessly compress each individual quantized feature $\bar{\F}_{t}^{s}$, using the compressed features $\hat{\F}_{t}^{s-1}$ from the coarser plane $s-1$ and $\hat{\F}_{t-1}^{s}$ from the previous time-step $t-1$ as spatial and temporal priors, respectively. 
Moreover, to capture long-term temporal dependencies, NVCC maintains a hidden temporal state that updates the temporal prior. By exploiting spatiotemporal correlations, NVCC generates spatiotemporal contexts to produce more accurate estimated distribution $P(\bar{\F}_{t}^{s} | \cdot)$, thereby enhancing contextual coding and lossless compression.

Meanwhile, we compress the canonical 3DGS. While previous studies have explored multi-rate 3DGS data compression~\cite{chen2024hac,wang2024contextgs} by employing neural contextual coding with anchor data structures to cluster 3D Gaussians, these approaches fall short for canonical 3DGS in 4DGS. This is because canonical 3DGS must store information covering entire dynamic scenes for world-to-canonical mapping with the 4D neural voxels~\cite{cao2023Hexplane}, leading to significant sparsity~\cite{wu20244d} that limits the effectiveness of anchor-based methods. Therefore, we adopt a compact 3DGS method that uses vector quantization~\cite{wang2024end, navaneet2024compgs, lee2024compact, papantonakis2024reducing} with an auxiliary quantized codebook to store spherical harmonic information and achieve lossy compression. 
However, previous works have overlooked redundancy within the codebook itself. To address this, we introduce a Vector Quantization Contextual Coding (VQCC) methodology that leverages an auxiliary hyperprior feature as context for further losslessly compressing those quantized vectors. To the best of our knowledge, this is the first approach to combine contextual coding with VQ strategies for compressing 3DGS data.


By integrating these two novel contextual coding techniques, we introduce the first 4DGS contextual coding framework, 4DGS-CC. Our framework can be optimized for multi-rate compression to meet diverse conditions while significantly reducing storage. Extensive experiments on dynamic scene compression benchmarks, including D-NeRF~\cite{pumarola2021d}, Neu3D~\cite{li2022neural}, and HyperNeRF~\cite{park2021hypernerf}, demonstrate that compared to our baseline 4DGS~\cite{wu20244d}, 4DGS-CC achieves an average storage reduction of approximately $12\times$ while maintaining comparable rendering quality. At the same time, we conducted further experiments with the Saro-GS\cite{yan20244d} method, thereby demonstrating the broader applicability of our approach. Our contributions are summarized as follows:

\begin{itemize}

\item A Gaussian deformation field compression method that disentangles 4D neural voxels into individual features and employs our newly proposed NVCC to exploit spatiotemporal redundancy.

\item A canonical 3DGS compression method, VQCC, which for the first time integrates vector quantization with a contextual coding strategy to reduce redundancy in both the sparse 3DGS data and its associated codebook.

\item A novel 4DGS compression framework, 4DGS-CC, which, to the best of our knowledge, is the first contextual coding method for 4DGS data that achieves multi-rate compression and produces the most compact dynamic scene representation using 4DGS-based methods.

\end{itemize}
\section{Related work}

\begin{figure*}
  \centering
    \includegraphics[width=1\linewidth]{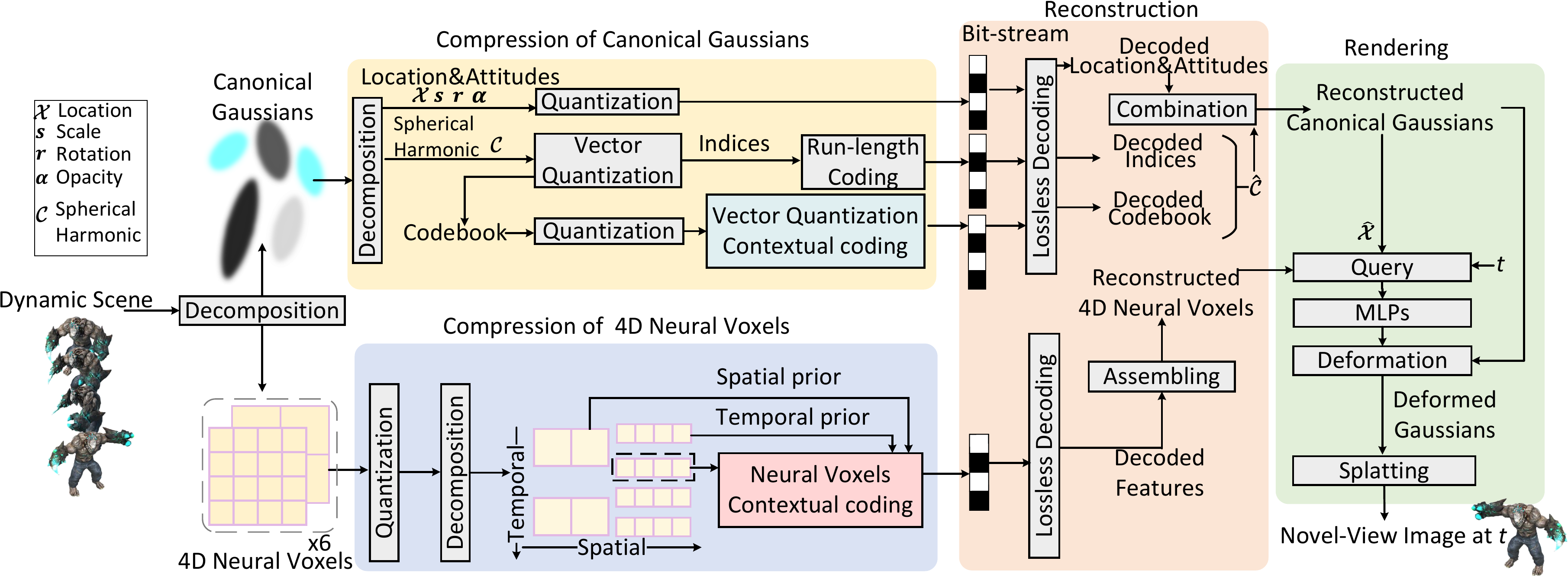}
    \vspace{-2mm}
    \caption{The overview of our 4DGS data compression framework. It consists of six main components, which are decomposition, compression of Canonical Gaussians, compression of 4D neural voxels, lossless decoding, reconstruction and rendering.}
    \vspace{-4mm}
    
  \label{fig:overview}
\end{figure*}
\noindent\textbf{Dynamic Scene Rendering.} It is a challenging task that creates new views of dynamic scenes from 2D images captured at different times. Recent 3D reconstruction techniques like NeRF and 3DGS have significantly advanced this field. Researchers extended NeRF to 4D for dynamic scene reconstruction~\cite{pumarola2021d, gan2023v4d, park2021hypernerf,fang2022fast}, and later methods decomposed 4D scenes into 2D planes for more compact representations~\cite{wu2024tetrirf, cao2023Hexplane, fridovich2023k}. However, these methods are limited by their rendering efficiency.

To address this issue, 3DGS rendering efficiency is utilized. Extending 3DGS to 4D involves creating 3D Gaussians at each timestamp, but this leads to high storage costs~\cite{luiten2024dynamic}. There are two main methods to extend 3DGS to 4D: explicit and implicit methods. Explicit methods manage the variation of Gaussian properties over time using functions like polynomial, Fourier, or learned functions~\cite{sun20243dgstream, li2024spacetime, katsumata2024compact}, or by adding a time dimension to 3D Gaussian primitives~\cite{yang2023real, duan20244d}. Implicit methods, also called deformable 3DGS~\cite{yang2024deformable, huang2024sc, wu20244d, xu2024grid4d, yan20244d}, separate the scene into a canonical 3D Gaussian set and a deformation field to explain temporal changes. Both methods offer good reconstruction, but explicit methods increase 3DGS primitive attributes, while implicit methods use a compact deformation field for better storage. 4DGS~\cite{wu20244d} is a deformable 3DGS method using a canonical Gaussian set and a deformation field with the Hexplane 4D neural voxels~\cite{cao2023Hexplane}, have demonstrated competitive results with relatively low storage requirements on both multi-view and monocular videos. Despite these improvements, storage of 4DGS data remains a critical bottleneck. To address this, we introduce the 4DGS-CC compression framework, which aims to further reduce storage costs while maintaining high rendering efficiency and reconstruction quality.

\noindent\textbf{Compact radiance field representations.} It have been an important research direction. For NeRF, compact grid~\cite{muller2022instant,cao2023Hexplane,fridovich2023k} structures have been shown to be very effective in reducing network size and improving accuracy, while exploiting contextual dependencies~\cite{chen2024far} between adjacent elements is also a key method for compressing NeRF. Various compression strategies have also been explored for 3D Gaussian fields (3DGS), including redundant Gaussian pruning~\cite{wang2024end,navaneet2024compgs,lee2024compact,papantonakis2024reducing}, spherical harmonic distillation or compression~\cite{fan2023lightgaussian,wang2024end}, vector quantization~\cite{wang2024end,navaneet2024compgs,lee2024compact,papantonakis2024reducing,fan2023lightgaussian}, entropy models, and more. Some approaches have represented Gaussian parameters as 2D grids~\cite{morgenstern2024compact} and applied image compression techniques.

Some works have also focused on developing compact dynamic 3DGS representations. For example, for monocular~\cite{kwak2025modecgsglobaltolocalmotiondecomposition} and multi-view videos~\cite{cho20244dscaffoldgaussiansplatting}, researchers have attempted to achieve compact dynamic 3DGS representations by leveraging the compact structure introduced in~\cite{lu2024scaffold}. However, these approaches do not constitute a multi-rate compression framework. At the same time, although~\cite{kwak2025modecgsglobaltolocalmotiondecomposition} focuses on compact deformable 3D Gaussian representations, it does not attempt to explore the redundancy inherent in 4D neural voxels.
Our method simultaneously accounts for the redundancy present in both the Gaussians and 4D neural voxels, and supports multi-rate compression.

\noindent\textbf{Neural Contextual Coding.}
In modern neural compression frameworks for images and videos, the ability to accurately predict probability distributions through sophisticated entropy modeling is of paramount importance. For instance, a notable study by Minnen et al.~\etal~\cite{minnen2018joint} devised an innovative compression strategy that fuses a hyperprior network with an autoregressive component, thereby capturing intricate spatial correlations within image data to substantially boost compression efficiency. This paradigm, which integrates multiple modeling techniques to better represent data redundancies, has been mirrored in several subsequent compression approaches~\cite{lu2019dvc,hu2021fvc,li2021deep,ChenIntra-Slice,han2024cra5,liu2023icmh,chen2023neural,hu2020improving,Chen_2022_CVPR,liu2025efficient,chen2024group}, further affirming the critical role that neural contextual coding play in advancing the state-of-the-art in data compression research.
Currently, a few works have attempted to apply neural contextual coding for compression in static scene representations, such as those in NeRF~\cite{chen2024far} and 3DGS~\cite{chen2024hac,wang2024contextgs,liu2024hemgs}, achieving promising results. However, these studies have not explored the application of neural contextual coding for compression in dynamic 3DGS scenarios. To the best of our knowledge, we are the first work to apply neural contextual coding techniques to the compression of Gaussian Splatting based dynamic scene representations.


\section{Methodology}

\vspace{-2mm}
\subsection{Preliminaries}
\vspace{-2mm}


In this study, we follow the deformable 3DGS methodology~\cite{wu20244d, yan20244d} to decompose 4DGS data into \textit{canonical 3DGS data} and a \textit{Gaussian deformation field}, which are subsequently compressed using \textit{Neural contextual Coding} techniques. In this section, we briefly review these three methodologies.

\noindent\textbf{Canonical 3DGS data.} 
In the deformable 3DGS framework, the canonical 3DGS data represents a sparse variant of the 3DGS data that encapsulates information covering the entire dynamic scene. Specifically,
3DGS is a technique for representing 3D scenes as point cloud, where each point is associated with a full 3D covariance matrix, $\Sigma$ centered at $\mathcal{X}$ to represent a Gaussian function $G(\mathcal{X})=e^{-\frac{1}{2} \mathcal{X}^T \Sigma^{-1} \mathcal{X}}$. 
To facilitate differentiable optimization, the covariance matrix $\Sigma$ is factorized as $\Sigma = R^T S^T S R$, where $S$ and $R$ are scaling and rotation matrix parameterized by the vector $s$ and $r$, respectively. 
Each Gaussian is assigned an opacity $\alpha$ and a view-dependent color, represented using spherical harmonics $\mathcal{C}$, which are used to compute the pixel values during rendering.
Hence, we will only use the location $\mathcal{X}$ and attributes $s$, $r$, $\alpha$, $\mathcal{C}$ to characterize the 3DGS data.

%

\label{Vector quantization Contextual coding}
\begin{figure}[t]
  \centering
   \includegraphics[width=1\linewidth]{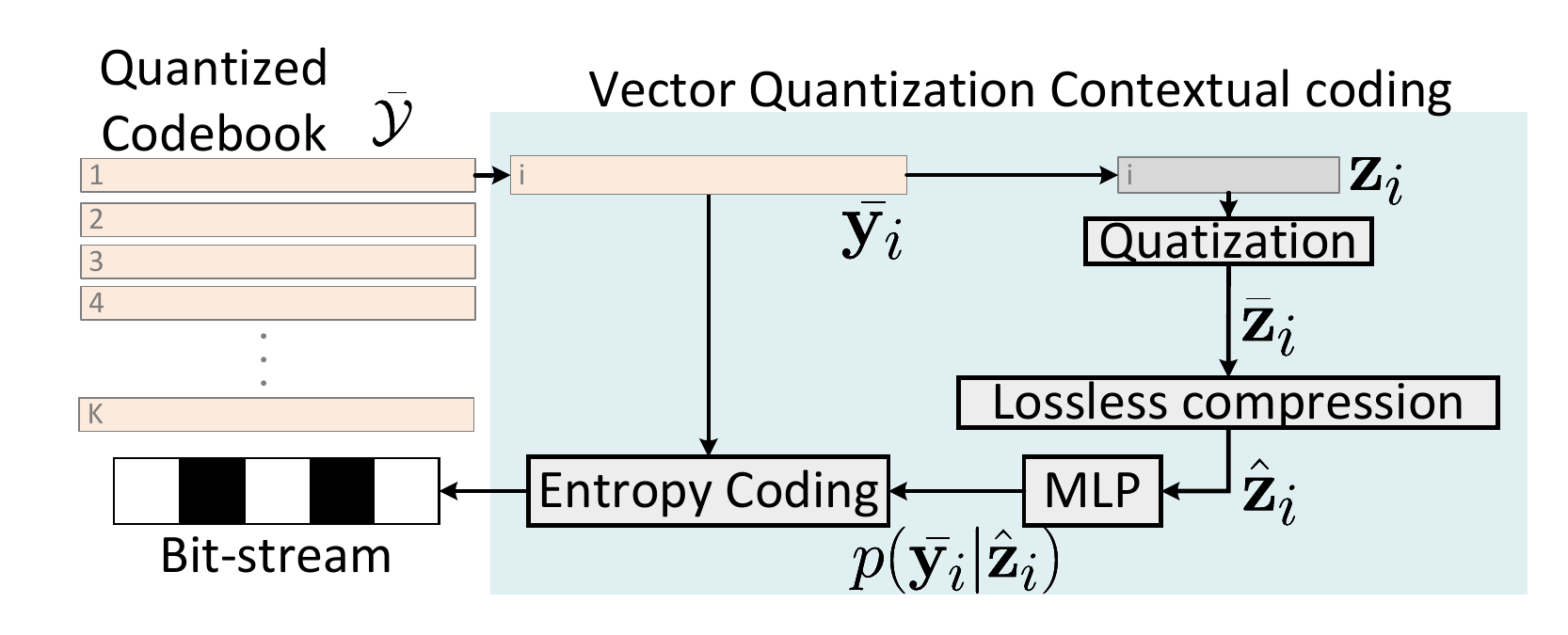}
\vspace{-3mm}
   \caption{Overview of Vector Quantization Contextual Coding.}
\vspace{-3mm}
    \label{fig:VCC}
\end{figure}
\label{Gaussian Deformation Field}
\noindent\textbf{Gaussian Deformation Field}. In our work, we primarily follow the framework presented in~\cite{wu20244d,yan20244d}, which employs 4D neural voxels to store spatiotemporal information for robust rendering in both monocular and multi-view scenarios.
Specifically, we employ the 4D neural voxels method described in~\cite{cao2023Hexplane}. For a given time-step \(t\), we concatenate \(t\) with the spatial coordinates $\mathcal{X}$ of each Gaussian and query the 4D neural voxels to retrieve the corresponding feature. This feature is then processed by the MLPs to generate the deformation parameters
$\{\Delta\mathcal{X}, \Delta{s}, \Delta{r}, \Delta\alpha, \Delta\mathcal{C}\}$ by using MLPs. 
The resulting deformation information is subsequently fused with the \textit{canonical 3DGS data} to produce deformed Gaussians, which are used to render the novel-view image at time-step $t$.




\label{Neural Contextual Coding}
\noindent\textbf{Neural Contextual Coding.}
We employ neural contextual coding techniques to losslessly compress a given quantized content $\bar{\A}$. According to information theory~\cite{shannon1948mathematical}, the cross entropy $H(q,p) = \mathbb{E}_{\bar{\A} \sim q} \left[-\log(p(\bar{\A}))\right] $ serves as a practical lower bound on the storage cost for losslessly compressing $\bar{\A}$. Here, $p(\bar{\A})$ denotes the estimated probability distribution of  $\bar{\A}$, while $q(\bar{\A})$, represents its true distribution. By refining $p(\bar{\A})$  to more closely approximate $q(\bar{\A})$, we can reduce $H(q,p)$, and consequently lower the storage cost. This procedure is known as entropy coding, where, in our work, the conditional probability $p(\bar{\A} | \cdot)$  is modeled based on previously compressed contents.

\vspace{-1mm}

\subsection{Overview}
\vspace{-1mm}
Here, we employ the deformable 3DGS framework as in 4DGS~\cite{wu20244d} to demonstrate our 4DGS-CC framework, as illustrated in Fig.~\ref{fig:overview}. It is worth noting that our methodology should be theoretically applicable to other deformable 3DGS frameworks. For instance, in our implementation, we also build upon the latest method SaRO-GS~\cite{yan20244d}.

\textbf{Decomposition.} We first decompose the dynamic scene into two primary components: Canonical Gaussians and a Gaussians deformation field (\ie, 4D neural voxels), each of which will be compressed separately.

\begin{figure}[t]
  \centering
   \includegraphics[width=1\linewidth]{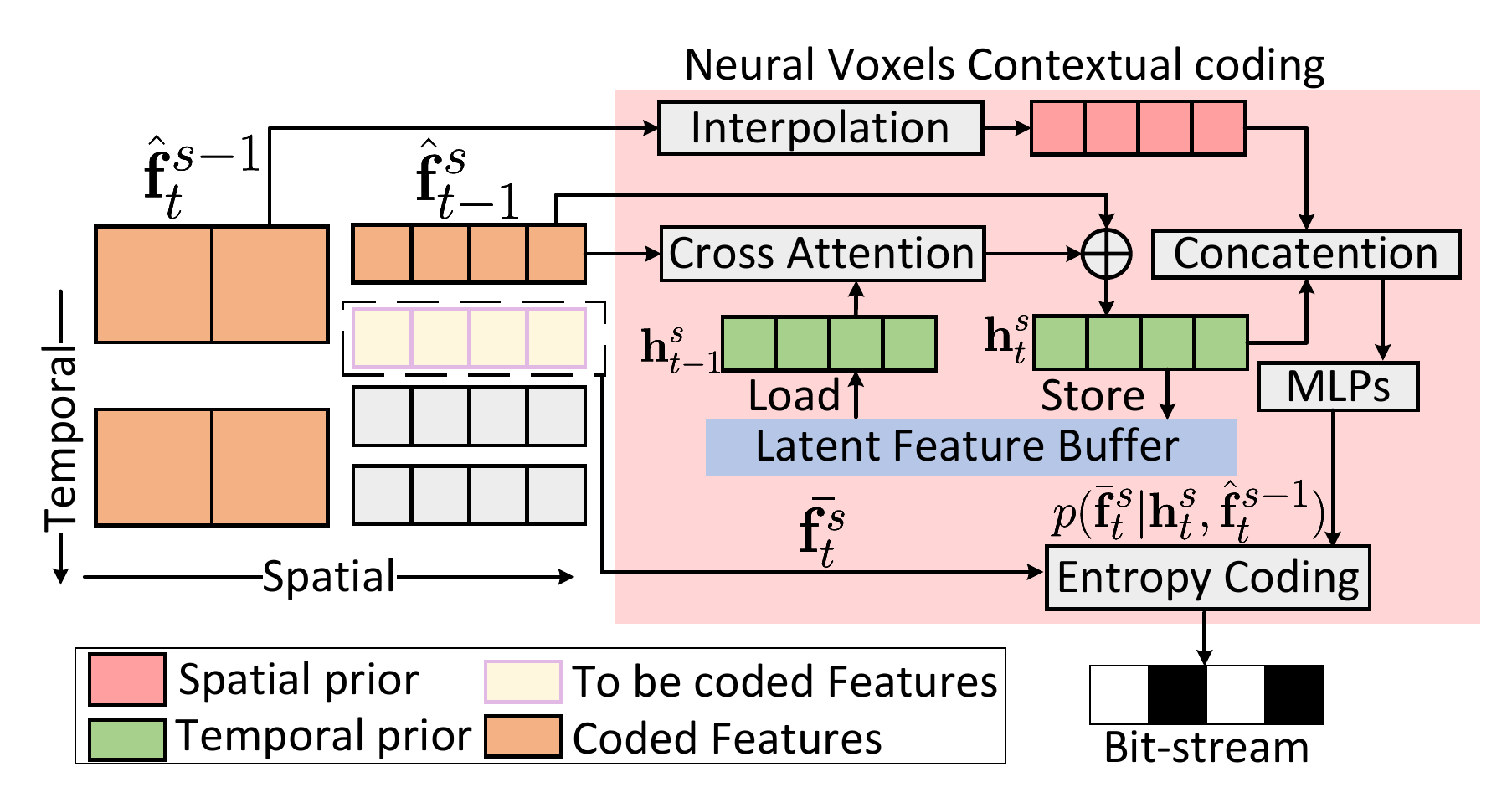}
\vspace{-3mm}
   \caption{Overview of Neural Voxels Contextual Coding.}
\vspace{-3mm}

    \label{fig:VCC}
\end{figure}
\textbf{Canonical Gaussian Compression.} The Canonical Gaussian data is characterized by the location $\mathcal{X}$ and attributes including scaling $s$, rotation $r$, and opacity $\alpha$, which will be compressed using a hybrid lossy–lossless compression scheme.
Specifically, $\mathcal{X}$ is first quantized at 16 bits and  $s, r$ and $\alpha$ quantized to 8 bits to achieve lossy compression, which will be losslessly stored directly.

For the lossy compression of the spherical harmonics $\mathcal{C}$,
we adopt a vector quantization strategy as described in~\cite{navaneet2024compgs}, which encodes the information into indices and a codebook. The indices are further losslessly compressed using a run-length coding algorithm into the bit-stream. Meanwhile, the codebook $\cY$ is first quantized into $\bar{\cY}$ , then is losslessly compressed using our VQCC strategy, which leverages a learned hyper-prior for lossless contextual coding. Detailed information about the VQCC is provided in \textit{Section~\ref{Vector quantization Contextual coding}}.

\textbf{4D Neural Voxel Compression.}
The 4D Neural Voxels are constructed as a set of multi-resolution Hexplanes. Their compression also follows a hybrid lossy–lossless scheme similar to the compression of Canonical Gaussian. Initially, we apply the quantization strategy proposed in~\cite{balle2018variational} directly to the 4D Neural Voxels for the lossy compression. Next, the quantized 4D Neural Voxels are decomposed into individual quantized latent features $\bar{\f}_t^s$, where the spatial dimension $s$ corresponds to the $s$-th level in a coarse-to-fine hierarchy, and the temporal dimension $t$ corresponds to the $t$-th row in each Hexplane storing the deformation information for a specific static scene at a given time-step inspired by~\cite{cao2023Hexplane}. 
To losslessly compress each individual quantized feature $\bar{\f}_t^s$, we introduce an innovative NVCC approach that leverages spatial and temporal priors, the compressed feature $\hat{\f}_{t}^{s-1}$ from the coarser plane $s-1$ and the compressed feature $\hat{\f}_{t-1}^{s}$ from the previous time-step $t-1$ as spatial and temporal priors, respectively. Detailed information on the NVCC approach is provided in \textit{Section~\ref{NVCC}}.

\textbf{Lossless Decoding}. 
During the decompression stage, all bit-streams are losslessly decoded. Here, we can obtain the decoded location $\mathcal{\hat{X}}$ and the decoded attributes $\hat{s}, \hat{r}, \hat{\alpha}$. To losslessly decode the codebook and feature  $\hat{\f}_t^s$, we employ an entropy decoding algorithm that reconstructs the vectors from the bit-stream using the estimated distributions produced by VQCC and NVCC, respectively. Detailed information is provided in the supplementary materials.

\textbf{Reconstruction.}
With compressed indices and codebook, we can assemble the reconstructed spherical harmonics $\mathcal{\hat{C}}$, which will be combined with compressed location $\mathcal{\hat{X}}$ and compressed attributes $\hat{s}, \hat{r}, \hat{\alpha}$ to obtain the reconstructed Canonical 3DGS data using the operations in~\cite{navaneet2024compgs}. Meanwhile, with those compressed features $\bar{\f}$, we can re-assemble the reconstructed 4D Neural Voxels. 

\textbf{Rendering.}
We use both reconstructed 4D Neural Voxels and Canonical 3DGS data to render the novel-view image at time-step $t$ using the standard operations as in \textit{Section~\ref{Gaussian Deformation Field}}.

\subsection{Vector quantization Contextual Coding}
\vspace{-1mm}

To losslessly compress the quantized codebook $\bar{\cY} = \{\bar{\y}_i\}$ we adopt a VQCC strategy as illustrated in Fig.~\ref{fig:VCC}. Inspired by the hyper-prior contextual coding techniques in neural image compression~\cite{balle2018variational}, we learn a hyper-prior feature $\z_i$ for each quantized code $\bar{\y}_i$. 
This hyper-prior feature is then quantized to produce $\bar{\z}_i$ and losslessly compressed using the uniform entropy coding algorithm as in~\cite{balle2016end} and subsequently losslessly compressed using the uniform entropy coding algorithm described in~\cite{balle2016end}, resulting in a compressed hyper-prior feature $\hat{\z}_i$. 
The compressed hyper-prior is used as prior information to estimate the distribution $p(\bar{\y}_i| \hat{\z}_i)$, where the distribution is parameterized by learned mean and variance vectors produced by MLPs, as detailed in~\cite{balle2018variational}. With the estimated distribution $p(\bar{\y}_i| \hat{\z}_i)$, we perform an entropy coding procedure outlined as in~\textit{Section~\ref{Neural Contextual Coding} on Neural Contextual Coding}, to losslessly transmit $\bar{y}_i$ into a bit-stream. To the best of our knowledge, this is the first work to investigate redundancy within the codebook by incorporating a hyper-prior in the 3DGS data compression research.

\subsection{Neural Voxels Contextual Coding}
\label{NVCC}
\vspace{-1mm}





To losslessly compress each quantized feature $\bar{\f}_t^s$ at $s$-th plane and $t$-th time-step (\ie, row), we introduce the NVCC methodology, which exploits the spatiotemporal redundancy inherent in the neural voxel. 
Similar to VQCC, we adopt a hyperprior contextual coding paradigm, where  $\bar{\f}_t^s$ is iteratively entropy coded using NVCC. Specifically, when compressing $\bar{\f}_t^s$, we utilize the compressed feature $\hat{\f}_{t}^{s-1}$ from the coarser plane $s-1$ as the spatial prior and the compressed feature $\hat{\f}_{t-1}^{s}$ from the previous time step $t-1$ as the temporal prior.

Using only the temporal prior $\hat{\f}_{t-1}^{s}$ at $t-1$ without proper alignment can introduce temporal disparity. To improve temporal consistency, we leverage long-term dependencies by incorporating more additional temporal priors, denoted as $\{\hat{\f}_{i}^{s}|i<t\}$. However, conditioning on all previously compressed temporal priors $\{\hat{\f}_{k}^{i}| i<t\}$ becomes impractical when dealing with a large number of time-step $t$. Inspired by~\cite{ChenIntra-Slice}, we adopt a recurrent architecture and introduce latent features $\h_t^s$ to summarize the accumulated information before time-step $t$. During our implementation, when coding $\bar{\f}_{t}^{s}$, we employ the following cross-attention mechanism to align the temporal prior  $\hat{\f}_{t-1}^{s}$ with the hidden feature $\h_{t-1}^s$ yielding a new hidden feature $\h_{t}^s$. This updated hidden feature will serve as an enhanced temporal prior for coding $\bar{\f}_t^s$ and will be iteratively fused with the compressed feature $\hat{\f}_{t}^{s}$ during the coding of $\bar{\f}_{t+1}^{s}$ at $t+1$. The final estimated distribution is defined as $p(\bar{\f}_t^s | \h_t^s, \hat{\f}_t^{s-1})$.

\vspace{-2mm}
\begin{equation}
\begin{split}
&\mathbf{q_t} = W_q \cdot \mathbf{h}_{t-1}^s,
\mathbf{k_t} = W_k \cdot \hat{\mathbf{f}}_{t-1}^{s},
\mathbf{v_t} = W_v \cdot \hat{\mathbf{f}}_{t-1}^{s},\\
&\mathbf{h}_{t}^s = \text{Softmax}\left(\frac{\mathbf{q_t} \cdot \mathbf{k_t}^T}{\sqrt{d}}\right) \cdot \mathbf{v_t} + \hat{\mathbf{f}}_{t-1}^{s}, 
\end{split}
\end{equation}
where $W_q,W_k,W_v$ are learnable projection matrices and $d$ represent the dimension of the vector $\mathbf{q}_t$.

Note that the first quantized feature $\bar{\f}_1^1$ at the first plane and the first time step is entropy coded using a uniform distribution $p(\bar{\f}_1^1)$ as in~\cite{balle2016end}. For the remaining features at the first plane, only temporal priors are available, and their estimated distribution is given by $p(\bar{\f}_t^1 | \h_{t}^1)$. In contrast, features at the first time step are encoded solely using spatial priors, yielding an estimated distribution $p(\bar{\f}_1^s|\hat{\f}_{t}^{s-1})$. Moreover, since we adopt the deformable 3DGS framework from~\cite{wu20244d,yan20244d}, which comprises six neural voxels and three of which are purely spatial by design, therefore the estimated distributions for these spatial-only voxels are produced exclusively using spatial priors.

\vspace{-1mm}
\subsection{Optimization}
\vspace{-1mm}
We optimize our 4DGS-CC framework by jointly minimizing both the original loss from the 4DGS baseline models and an additional storage-related loss, as detailed below.
\begin{equation}
\mathcal{L} = \mathcal{L}_{\mathrm{4DGS}} + \lambda_e \mathcal{L}_{\mathrm{voxels}} + \lambda_c \mathcal{L}_{\mathrm{code}}.
\end{equation}

Here, \(\mathcal{L}_{\mathrm{4DGS}}\) denotes the original 4DGS loss~\cite{wu20244d,yan20244d}, which primarily comprises a rendering optimization term that quantifies the discrepancy between the rendered and ground-truth images, along with several regularization terms to enhance the quality. Additionally, $\mathcal{L}_{\mathrm{code}}$ and $\mathcal{L}_{\mathrm{voxels}}$ represent the storage costs for the compressed codebook and neural voxels, respectively. We optimize these costs by directly minimizing the cross-entropy values produced by the VQCC and NVCC modules, thereby enabling differentiable optimization following NIC methods~\cite{balle2018variational,balle2016end,minnen2018joint,1,6}. The user-specific hyperparameters $\lambda_e$ and $\lambda_c$ balance these components to achieve an optimal storage-distortion trade-off under varying storage requirements.



\begin{table}[h]
\centering
\footnotesize
\setlength{\tabcolsep}{3pt}
\renewcommand{\arraystretch}{1.3}

\vspace{-2mm}
\begin{tabular}{lcccc}
\noalign{\hrule height 1pt}
\rowcolor[HTML]{FFFFFF} 
\textbf{Methods} & \textbf{PSNR$\uparrow$} & \textbf{SSIM$\uparrow$} & \textbf{LPIPS$\downarrow$} & \textbf{Size$\downarrow$} \\
\noalign{\hrule height 1pt}
TiNeuVox-B~\cite{fang2022fast}       & 32.67 & 0.97  & 0.040      & 48    \\
Hexplane~\cite{cao2023Hexplane}        & 31.04 & 0.97  & 0.040      & 38    \\
V4D~\cite{gan2023v4d}              & 33.72 & 0.98  & 0.020      & 377   \\
Kplanes~\cite{fridovich2023k}          & 31.61 & 0.97  & -         & 418   \\ \hline
4DGS~\cite{wu20244d}             & 34.07 &   \cellcolor{pink}{0.979} & 0.026     & 23.3 \\
Saro-GS~\cite{yan20244d}          & \cellcolor{pink}{35.04} & 0.977 & \cellcolor{pink}{0.025}     & 49.9 \\
C-D3DGS~\cite{katsumata2024compact}          & 32.19 & -     & 0.040      & 159 \\
Real-time 4DGS~\cite{yang2023real}   & 31.32 & 0.961 & 0.043     & 1068
\\ \hline
Ours (4DGS) high    & 34.12 & \cellcolor{yellow}{0.978} & 0.028     & 1.9   \\
Ours (4DGS) mid     & 33.99 & 0.977 & 0.029     & \cellcolor{yellow}{1.8}   \\
Ours (4DGS) low     & 33.94 & 0.977 & 0.030      & \cellcolor{pink}{1.7}   \\
Ours (Saro-GS) high & \cellcolor{yellow}{34.9}  & 0.976 & \cellcolor{yellow}{0.026}     & 5.6   \\
Ours (Saro-GS) mid  & 34.79  & 0.976 & 0.027     & 4.9 \\
Ours (Saro-GS) low  & 34.69 & 0.976 & 0.028     & 4.4  \\
\noalign{\hrule height 1pt}
\end{tabular}
\vspace{-1mm}
\caption{Quantitative comparison results on D-NeRF dataset. The best and second-best results are highlighted in \colorbox{pink}{\textcolor{black}{red}} and \colorbox{yellow!50}{\textcolor{black}{yellow}} cells, respectively. Size values are measured in megabytes (MB). At three distinct storage levels, our 4DGS variant achieves compression ratios of $\textbf{12.3}\times$, $\textbf{12.9}\times$, and $\textbf{13.7}\times$ , while our Saro-GS variant achieves compression ratios of $\textbf{8.9}\times$,$\textbf{10.2}\times$,$\textbf{11.3}\times$, compared to baseline methods 4DGS and Saro-GS.}
\label{d-nerf}
\vspace{-2mm}

\end{table}

\vspace{-1mm}
\section{Experiment}
\vspace{-1mm}



\subsection{Experiment Protocols}
\noindent\textbf{Datasets.}  We conducted comprehensive experiments on two monocular datasets, D-NeRF~\cite{pumarola2021d} and HyperNeRF~\cite{park2021hypernerf}, and one multi-view dataset, Neu3D~\cite{li2022neural}. Specifically, the monocular dataset, D-NeRF, is a synthetic monocular video dataset specifically designed for monocular settings, where camera poses are generated in a near-random manner for each timestamp. The monocular dataset, HyperNeRF, uses points computed by structure-from-motion (SfM) from 200 randomly selected frames, where we selected the vrig subset comprising four scenes. The multi-view dataset is captured using 15 to 20 static cameras, covers longer durations, and exhibits complex camera motions.
The rendering resolutions were set to 800 × 800 for D-NeRF, 536 × 900 for HyperNeRF, and 1352 × 1024 for Neu3D.
\begin{table}[h]
\centering
\footnotesize
\setlength{\tabcolsep}{3pt}
\renewcommand{\arraystretch}{1.3}

\vspace{-2mm}
\begin{tabular}{lcccc}
\noalign{\hrule height 1pt}
\rowcolor[HTML]{FFFFFF} 
\textbf{Methods} & \textbf{PSNR$\uparrow$} & \textbf{DSSIM$\downarrow$} & \textbf{LPIPS$\downarrow$} & \textbf{Size$\downarrow$} \\
\noalign{\hrule height 1pt}
Hexplane~\cite{cao2023Hexplane}        & 31.71 & \cellcolor{pink}{0.014}  & 0.075      & 200  \\
Kplanes~\cite{fridovich2023k}         & 31.63 & 0.018  &            & 311  \\ \hline
 4DGS~\cite{wu20244d}            & 31.49 & 0.0151 & 0.058      & 38.1 \\
Saro-GS~\cite{yan20244d}         & \cellcolor{pink}{32.03} & \cellcolor{yellow}{0.0142} & \cellcolor{pink}{0.044}      & 385  \\
C-D3DGS~\cite{katsumata2024compact}          & 30.46 & -      & -          & 338  \\
Real-time 4DGS~\cite{yang2023real}   & 32.01 &  \cellcolor{pink}{0.014}  & 0.055      & 6270 \\
Dynamic 3DGS~\cite{luiten2024dynamic}     & 30.46 & 0.019  & 0.099      & 2772 
\\ \hline
Ours (4DGS) high    & 31.54 & 0.0155 & 0.062       & 3.2  \\
Ours (4DGS) mid    & 31.34 & 0.0155 & 0.064       & \cellcolor{yellow}{2.6}  \\
Ours (4DGS) low     & 30.97 & 0.0160  & 0.065       & \cellcolor{pink}{2.3}  \\
Ours (Saro-GS) high & \cellcolor{yellow}{31.93} & 0.0147 & \cellcolor{yellow}{0.045}      & 20.9 \\
Ours (Saro-GS) mid  & 31.73 & 0.0153 & 0.048      & 18.5 \\
Ours (Saro-GS) low  & 31.60  & 0.0154 & 0.047      & 17.0
\\\hline
\noalign{\hrule height 1pt}

\vspace{-1mm}
\end{tabular}
\vspace{-2mm}

\caption{Quantitative comparisons results on Neu3D dataset. The best and second-best results are highlighted in \colorbox{pink}{\textcolor{black}{red}} and \colorbox{yellow!50}{\textcolor{black}{yellow}} cells, respectively. The size values are measured in megabytes (MB). At three distinct storage levels, our 4DGS variant achieves compression ratios of \textbf{$\textbf{11.9}\times$}, \textbf{$\textbf{14.6}\times$}, and \textbf{$\textbf{16.6}\times$} , while our Saro-GS variant achieves compression ratios of \textbf{$\textbf{18.4}\times$},{$\textbf{20.8}\times$},{$\textbf{22.6}\times$}, compared to baseline methods 4DGS and Saro-GS.
}
\label{neu3d}

\end{table}

\noindent\textbf{Evaluation Metric.}
We measure the storage used for 4DGS data using the metric of storage size in megabytes (MB). The quality of the rendered images from the 4DGS methods is evaluated using peak-signal-to-noise ratio (PSNR), perceptual quality measure LPIPS \cite{zhang2018unreasonable}, structural similarity index (SSIM) \cite{wang2004image} and its extensions including structural dissimilarity index measure DSSIM.

\noindent\textbf{Implementation Details.}
Although our 4DGS-CC framework can theoretically be applied to most deformable 3DGS dynamic scene methods, we have chosen to implement it on two representative approaches: 4DGS~\cite{wu20244d} (referred to as Ours (4DGS)) and the latest work SaRO-GS~\cite{yan20244d} (referred to as Ours (SaRO-GS)). Both methods support multi-view and monocular settings.
We inherit the hyperparameter configurations (\eg, learning rate for the Gaussian's attitude and the Gaussian deformation field) from~\cite{wu20244d, yan20244d}. To achieve multi-rate compression, we simply use both the codebook size and the user-specific hyperparameter $\lambda$ to adjust storage, denoting the resulting configurations as high, medium, and low storage rates. Specific hyperparameter settings are provided in the supplementary materials. We emphasize that further tuning of the codebook size and $\lambda$, can yield additional storage rate options.
We optimized our framework using the Adam optimizer and implemented it in PyTorch with CUDA support. Our experiments based on 4DGS (\ie, Ours (4DGS)) were conducted on an NVIDIA 3090 GPU, while those based on Saro-GS (\ie, Ours (Saro-GS)) were executed on an NVIDIA A100 GPU.

\begin{figure*}
  \centering
    \includegraphics[width=1\linewidth]{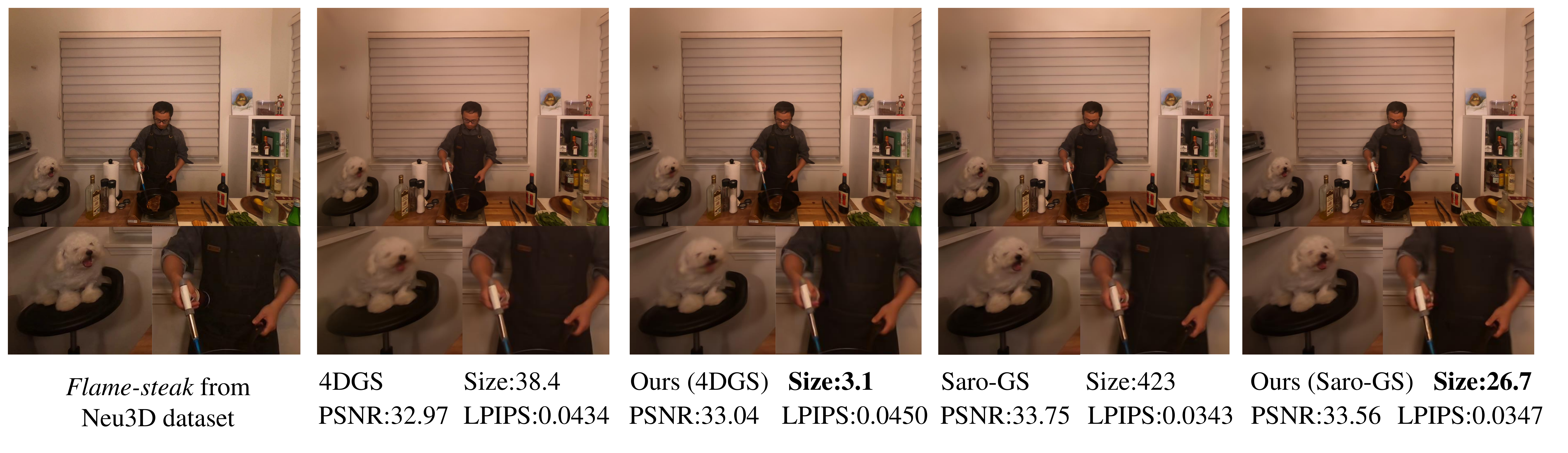}
    \caption{Visualization comparison between our baseline and our methods  on the ``Flame-steak" scene from the Neu3D dataset. PSNR, LPIPS and the size (MB) of the scene are reported.}
    \label{Vis}
\end{figure*}

\noindent\textbf{Benchmarks.} 
We compare our 4DGS-CC framework with state-of-the-art dynamic scene reconstruction methods. First, we include NeRF-based dynamic scene methods which utilize 4D neural voxels, which are Hexplane~\cite{cao2023Hexplane}, V4D~\cite{gan2023v4d}, TiNeuVox~\cite{fang2022fast}, and Kplanes~\cite{fridovich2023k}. We also include 3DGS dynamic scene rendering approaches, namely, Dynamic 3DGS~\cite{luiten2024dynamic}, C-D3DGS~\cite{katsumata2024compact}, Deformable 3DGS~\cite{yang2024deformable} and Real-time 4DGS~\cite{yang2023real}, along with our baselines Saro-GS~\cite{yan20244d} and 4DGS~\cite{wu20244d}. It is important to note that some methods cannot be straightforwardly applied to both viewing settings (\ie,  monocular and multi-view). For example, Dynamic 3DGS \cite{luiten2024dynamic} is applicable only in a multi-view setting, not in a monocular setting.

\begin{table}[h]
\centering
\footnotesize
\setlength{\tabcolsep}{3pt}
\renewcommand{\arraystretch}{1.3}

\vspace{-2mm}
\begin{tabular}{lcccc}
\noalign{\hrule height 1pt}
\rowcolor[HTML]{FFFFFF} 
\textbf{Methods} & \textbf{PSNR$\uparrow$} & \textbf{SSIM$\uparrow$} & \textbf{LPIPS$\downarrow$} & \textbf{Size$\downarrow$} \\
\noalign{\hrule height 1pt}
TiNeuVox-B~\cite{fang2022fast} & 24.3 & - & - & 48  \\
V4D~\cite{gan2023v4d}        & 24.8 & - & - & 377 \\ \hline
4DGS~\cite{wu20244d}            & 25.17 & \cellcolor{pink}{0.685} & \cellcolor{pink}{0.33}  & 65   \\
C-D3DGS~\cite{luiten2024dynamic}         & \cellcolor{pink}25.6  & -     & -     & 720  \\
Deformable 3DGS~\cite{yang2024deformable} & 22.72 & 0.616 & 0.35  & 138  \\ \hline
Ours (4DGS) high   & \cellcolor{yellow}{25.21} & \cellcolor{yellow}{0.683} & \cellcolor{yellow}{0.34}  & 5.0    \\
Ours (4DGS) mid     & 25.18 & 0.682 & 0.34  & \cellcolor{yellow}4.4 \\
Ours (4DGS) low    & 25.11 & 0.678 & 0.35  & \cellcolor{pink}{3.8} \\
\noalign{\hrule height 1pt}
\end{tabular}
\caption{Quantitative comparisons result on the HyperNeRF dataset. The best and second-best results are highlighted in \colorbox{pink}{\textcolor{black}{red}} and \colorbox{yellow!50}{\textcolor{black}{yellow}} cells, respectively. The size values are measured in megabytes (MB). At three distinct storage levels, our 4DGS variant achieves compression ratios of {$\textbf{13.0}\times$}, {$\textbf{14.7}\times$} and {$\textbf{17.8}\times$} compared to baseline methods 4DGS.
}
\label{hypernerf}
\vspace{-2mm}
\end{table}

\subsection{Experimental Results}
We present quantitative comparison results on the D-NeRF~\cite{pumarola2021d}, Neu3D~\cite{li2022neural}, and HyperNeRF datasets~\cite{park2021hypernerf} in Table \ref{d-nerf}, Table \ref{neu3d}, and Table \ref{hypernerf}, respectively. Notably, our 4DGS-based variant with a high storage-rate setting (\ie, Ours (4DGS) high) achieves even better reconstruction performance than the baseline 4DGS method, while obtaining compression ratios of 12.3×, 11.9×, and 13.0× on the D-NeRF~\cite{pumarola2021d}, Neu3D~\cite{li2022neural}, and HyperNeRF datasets~\cite{park2021hypernerf} datasets, respectively.

Similarly, our Saro-GS-based variant with a high storage-rate setting (\ie, Ours (Saro-GS) high) achieves comparable reconstruction performance to the baseline Saro-GS method, while delivering compression ratios of 8.9× and 18.4× on the D-NeRF~\cite{pumarola2021d} and Neu3D~\cite{li2022neural} datasets. Additionally, our Saro-GS-based variant with a low storage-rate setting (\ie, Ours (Saro-GS) low) achieves compression ratios of up to 11.3× and 22.6× on the D-NeRF~\cite{pumarola2021d} and Neu3D~\cite{li2022neural} datasets, respectively, with only a slight reduction in reconstruction quality. This indicates that our methods accommodate a wide range of compression scenarios, from low to high qualit, to meet various application needs.

\subsection{Ablation Study}

\begin{figure}
  \centering
   \includegraphics[width=0.85\linewidth]{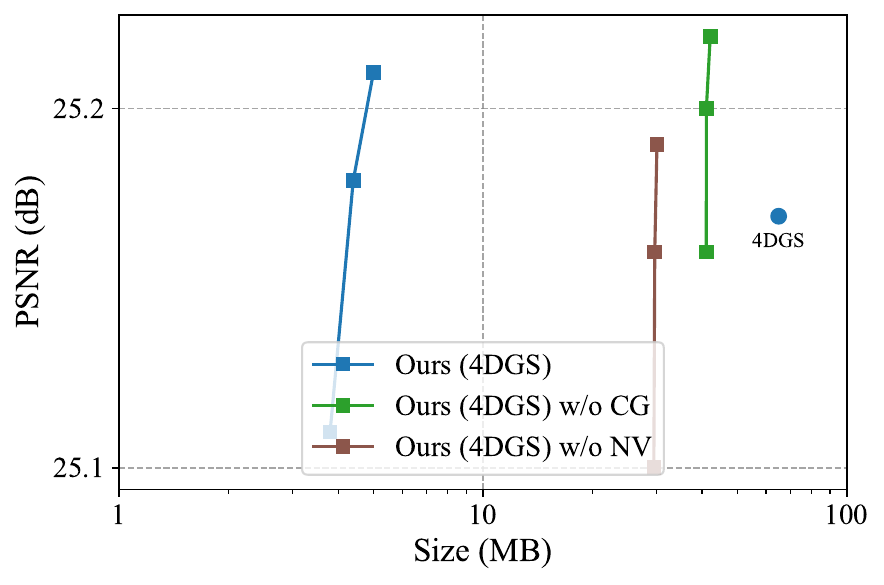}
    \vspace{-2mm}
   \caption{Ablation study of two compression components on the HyperNeRF dataset.
   (1) \textbf{4DGS:} Our baseline 4DGS.
   (2) \textbf{Ours (4DGS) w/o CG:} Our method based on 4DGS without the compression of Canonical Gaussian.
   (3) \textbf{Ours (4DGS) w/o NV:} Our method based on 4DGS without the compression of Neural Voxel.
   (4) \textbf{Ours (4DGS):} Our method based on 4DGS.}
    \label{fig:ab1}
\end{figure}

\textbf{Effectiveness of Individual Compression Components.} We conduct extensive ablation studies on our 4DGS-based method (\ie, Ours (4DGS)) to evaluate the impact of compressing two components, which are Canonical Gaussians and 4D Neural Voxels, on the HyperNeRF dataset. Specifically, we examine two variants: Ours (4DGS) w/o CG, which compresses only the 4D Neural Voxels, and Ours (4DGS) w/o NV, which compresses only the Canonical Gaussians. As shown in Fig.~\ref{fig:ab1}, compressing either component alone yields significant storage savings compared to the baseline 4DGS. Moreover, combining these two components in our full 4DGS-based method (\ie, Ours (4DGS)) achieves further storage reductions of 89.0\% and 86.2\% compared to the variants, Ours (4DGS) w/o CG and Ours (4DGS) w/o NV, respectively. This observation indicates that simultaneously compressing both components in a dual-compression fashion is highly beneficial.

\begin{figure}
  \centering
   \includegraphics[width=0.85\linewidth]{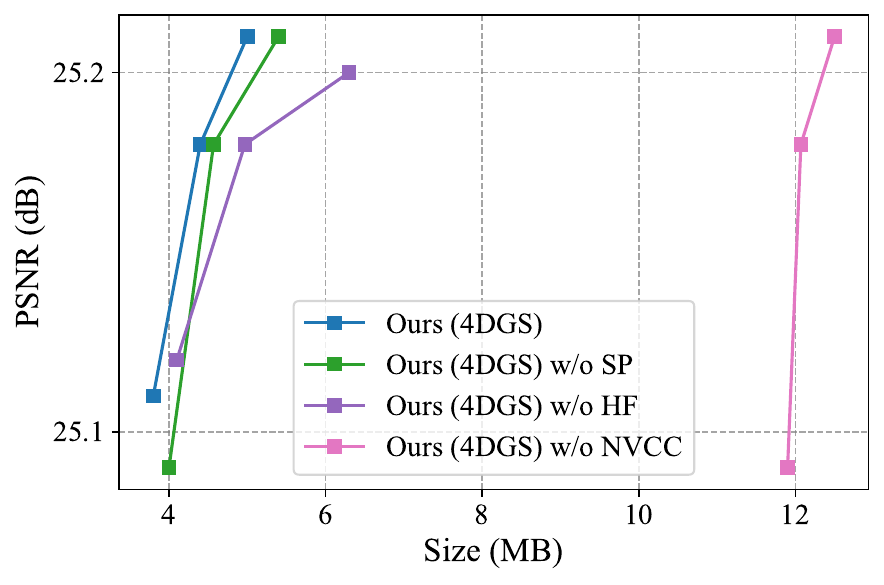}
    \vspace{-2mm}
   \caption{Ablation study of components within NVCC strategy on the HyperNeRF dataset 
   (1) \textbf{Ours (4DGS) w/o NVCC:} Our method based on 4DGS without using NVCC.
   (2) \textbf{Ours (4DGS) w/o SP:} Our method based on 4DGS without using spatial prior when performing NVCC.
   (3) \textbf{Ours (4DGS) w/o HF:} Our method based on 4DGS without updating the temporal priors with the hidden feature when performing NVCC.
   (4) \textbf{Ours (4DGS):} Our method based on 4DGS.}
    \label{fig:ab2}
    \vspace{-4mm}
\end{figure}

\textbf{Effectiveness of each compoents in NVCC.}
We also conduct ablation studies on our 4DGS-based method (\ie, Ours (4DGS)) on the HyperNeRF dataset to evaluate the effectiveness of each individual component within NVCC when losslessly compressing the quantized features of 4D Neural Voxels. Specifically, we examine three variants: the first variant (\ie, Ours (4DGS) w/o SP) uses only temporal priors, omitting the spatial prior; the second variant (\ie Ours (4DGS) w/o HF) does not employ the hidden feature updating mechanism for temporal priors; and the third variant (Ours (4DGS) w/o NVCC) does not apply NVCC for further lossless compression of the quantized 4D Neural Voxels. As shown in Fig.~\ref{NVCC}, both the Ours (4DGS) w/o HF and Ours (4DGS) w/o SP variants, which incorporate NVCC components, achieve significant storage savings compared to the variant that does not use NVCC, saving 63\% and 65\% respectively. Furthermore, combining all components leads to further performance enhancements. This observation underscores the importance of these components in the lossless compression of 4D Neural Voxels.

\textbf{Effectiveness of VQCC.} We further conduct ablation studies on HyperNeRF~\cite{park2021hypernerf} dataset for our 4DGS-based method (\ie, Ours (4DGS)) to evaluate the effectiveness of VQCC, which losslessly compresses the quantized codebook. In our experiments, we directly store the quantized codebook when compressing the Canonical Gaussians without applying VQCC (\ie, Ours (4DGS) w/o VQCC). It is observed that this variant introduces an extra storage cost, demonstrating the importance of effective lossless compression of the codebook.

\begin{figure}
  \centering
   \includegraphics[width=0.85\linewidth]{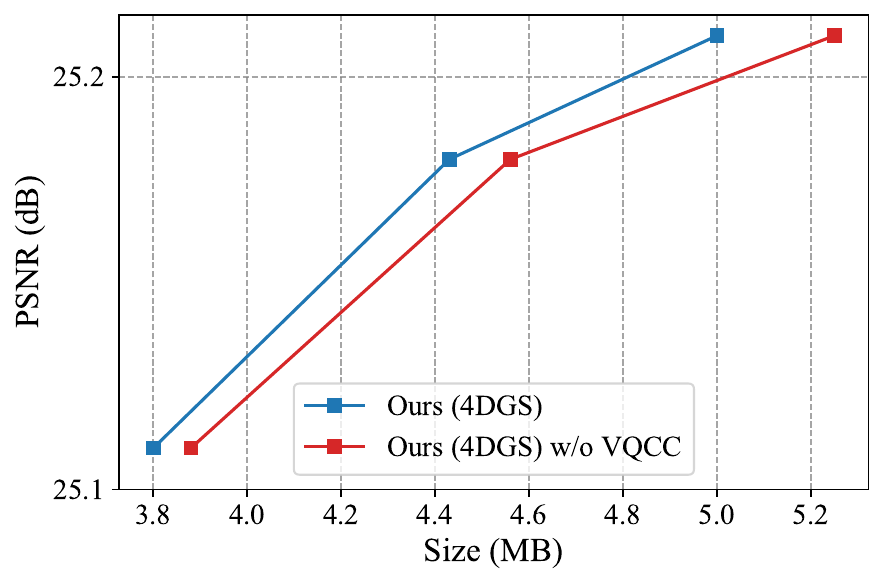}
    \vspace{-2mm}
   \caption{Ablation study of VQCC strategy on the HyperNeRF dataset.
   (1) \textbf{Ours (4DGS) w/o VQCC:} Our method based on 4DGS without using VQCC to further losslessly compress the codebook.
   (2) \textbf{Ours (4DGS):} Our method based on 4DGS.}
   
    \label{fig:ab3}
\end{figure}
\subsection{Visualization}
Our visualization results are displayed in Fig. \ref{Vis}, showing rendered outputs for the ``Flame-steak" scene from the Neu3D dataset~\cite{li2022neural} dataset.
It can be observed that even under high compression ratios, our approach preserves the rendering fidelity of the original method in challenging regions. For example, in the ``Flame-steak" scene, both our method and Saro-GS~\cite{yan20244d} capture details of the face of the dog.The maintained or improved fidelity metrics (\eg, PSNR and LPIPS) further demonstrate the superior quality of our method.


\section{Conclusion}
In this work, we propose the first neural contextual coding framework for compressing 4DGS data. Our approach builds on the deformable 3DGS framework and performs a lossy-lossless compression of two key components: Canonical Gaussians and 4D Neural Voxels. After applying lossy compression to these components, we further enhance lossless compression by introducing two innovative neural contextual coding strategies, VQCC and NVCC to entropy coding.
Extensive experiments demonstrate that our framework achieves significant storage reductions compared to other 4DGS methodologies while enabling variable-rate compression. We believe this work bridges the gap between 4DGS and data compression research, paving the way for broader future investigations.

\noindent\textbf{Limitations and Future Work.} Currently, our 4DGS-CC framework is implemented only with deformable 3DGS approaches that utilize Hexplane 4D Neural Voxels in the deformation field. In future work, we plan to extend our framework to a broader range of deformable 3DGS methods, including those that employ 4D Hash Grids in the deformation field, thereby enhancing its general applicability.

\small
\bibliographystyle{ieeenat_fullname}
\bibliography{main}
\columnbreak
\section{Appendix}
This section will provide details about decoding process in~\cref{sec:decode}, hyperparameter configuration in~\cref{sec:hyperparameter}, quantitative results of each scene in~\cref{sec:result}, the RD curve of our reported result~\cref{sec:rd-curve}, 

\subsection{Lossless Entropy Decoding}

\label{sec:decode}
We can directly adopt an entropy decoding algorithm to recover the quantized contents from the stored bitstream by using the estimated conditional distributions produced by our VQCC and NVCC modules, respectively. The decoding process mirrors the encoding process. Taking VQCC as an example, to obtain the decoded feature $\hat{\y}_i$, we first receive its bitstream. Then, based on the compressed hyperprior $\hat{\z}_i$, we estimate the distribution $p(\hat{\y}_i | \hat{\z}_i)$ and perform lossless entropy decoding, which reconstructs the feature $\hat{\y}_i$. This decoded feature is identical to the original quantized feature $\bar{\y}_i$, ensuring that no information distortion occurs.

\subsection{Hyperparameter settings}
\label{sec:hyperparameter}
We set $\lambda_c$ in Eq. (2) to 0.01, and the values of $\lambda_e$ and the codebook size are summarized in Table~\ref{table:codebook}.

\subsection{Quantitative Results of Each Scene}
\label{sec:result}
We report per-scene compression performance across the D-NeRF, Neu3D, and HyperNeRF datasets.
Specifically, for the D-NeRF dataset, we present the per-scene performance of our method based on 4DGS~\cite{wu20244d} and Saro-GS~\cite{yan20244d} in Tables~\ref{table:dnerf1} and~\ref{table:dnerf2}.
For the Neu3D dataset, the per-scene performance based on 4DGS~\cite{wu20244d} and Saro-GS~\cite{yan20244d} is shown in Tables~\ref{table:dynerf1} and~\ref{table:dynerf2}.
For the HyperNeRF dataset, we provide the per-scene performance of our method based solely on 4DGS~\cite{wu20244d}, reported in Table~\ref{table:hypernerf1}.

\subsection{RD Curve of our result}
\label{sec:rd-curve}

\begin{figure*}[t]
  \centering
   \includegraphics[width=0.7\linewidth]{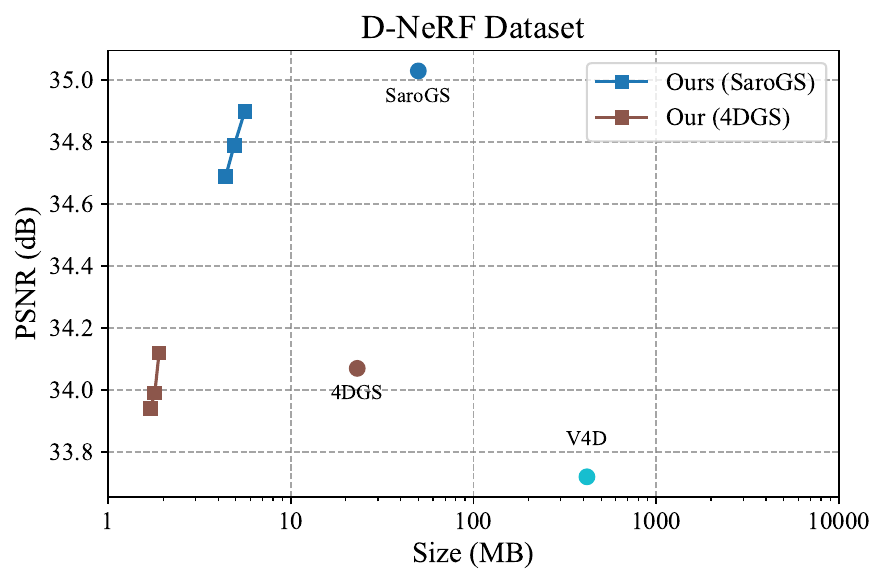}

   \caption{RD Curve on D-NeRF Dataset.}
    \label{fig:Dnerf}
\end{figure*}

\begin{figure*}[t]
  \centering
   \includegraphics[width=0.7\linewidth]{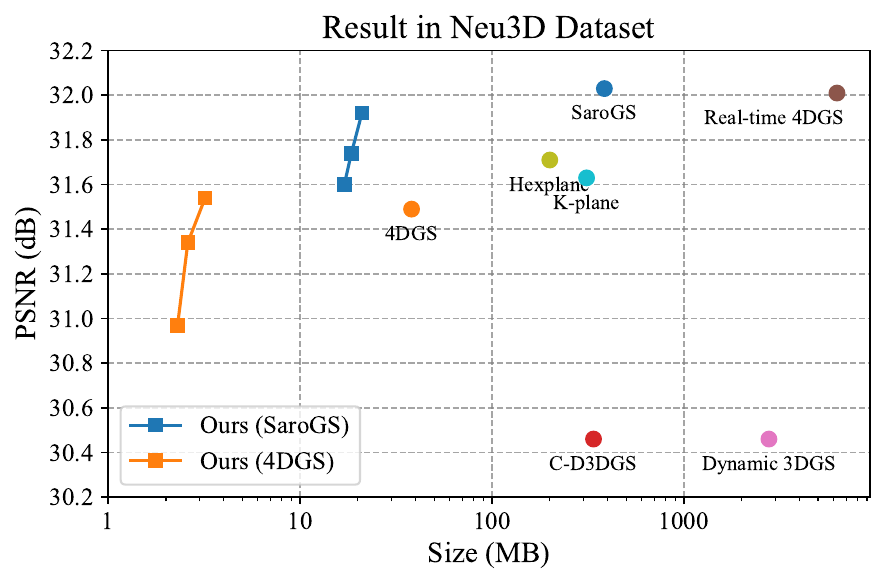}

   \caption{RD Curve on Neu3D Dataset.}
    \label{fig:Dynerf}
\end{figure*}

We present the Rate-Distortion curves for the D-NeRF dataset in Fig.~\ref{fig:Dnerf} and for the Neu3D dataset in Fig.~\ref{fig:Dynerf}.

\begin{table}[htbp]
  \centering

  \begin{tabular}{llcc}
  
    \toprule
    Method    & Dataset   & \(\lambda_e\)       & Codebook Size \\
    \midrule
    \multirow{3}{*}{4DGS} 
              &     & \(1\times10^{-3}\)  & 4096 \\
              & D-NeRF          & \(1\times10^{-4}\)  & 6144 \\
              &           & \(1\times10^{-5}\)  & 8192 \\
    \midrule
    \multirow{3}{*}{4DGS} 
              &    & \(1\times10^{-3}\)  & 4096 \\
              &  Neu3D         & \(1\times10^{-4}\)  & 8192 \\
              &          & \(1\times10^{-5}\)  & 16384 \\
    \midrule
    \multirow{3}{*}{4DGS} 
              &  & \(1\times10^{-3}\)  & 4096 \\
              &  HyperNeRF         & \(1\times10^{-4}\)  & 8192 \\
              &           & \(1\times10^{-5}\)  & 16384 \\
    \midrule
    \multirow{3}{*}{Saro-GS} 
              &     & \(1\times10^{-3}\)  & 8192 \\
              & D-NeRF          & \(1\times10^{-4}\)  & 16384 \\
              &           & \(1\times10^{-5}\)  & 32768 \\
    \midrule
    \multirow{3}{*}{Saro-GS} 
              &     & \(1\times10^{-3}\)  & 32768 \\
              & Neu3D         & \(1\times10^{-4}\)  & 65536 \\
              &          & \(1\times10^{-5}\)  & 140000 \\
    \bottomrule
  \end{tabular}
    \caption{Our hyperparameter settings across different methods and datasets, listed from top to bottom, correspond to increasing bit rates (from low to high).}
  \label{table:codebook}
    
\end{table}

\begin{table}[h]
\begin{tabular}{llcccc}
\hline
Rate       & Name               & PSNR  & SSIM  & LPIPS  & Size \\
\hline
\multirow{4}{*}{High} 
&vrig-chicken & 28.78 & 0.817 & 0.28  & 3.05 \\
&3dprinter    & 21.99 & 0.703 & 0.315 & 5.03 \\
&broom2       & 22.04 & 0.363 & 0.562 & 5.21 \\
&peel-banana  & 28.01 & 0.851 & 0.201 & 6.73 \\
\hline
\multirow{4}{*}{Mid} 
&vrig-chicken & 28.72 & 0.816 & 0.28  & 2.5   \\
&3dprinter    & 21.99 & 0.701 & 0.325 & 4.35  \\
&broom2       & 21.99 & 0.362 & 0.564 & 4.675 \\
&peel-banana  & 28.03 & 0.85  & 0.204 & 6.22 \\
\hline
\multirow{4}{*}{Low} 
&vrig-chicken & 28.71 & 0.815 & 0.289 & 2.43 \\
&3dprinter    & 21.96 & 0.701 & 0.323 & 3.67 \\
&broom2       & 21.9  & 0.352 & 0.586 & 3.98 \\
&peel-banana  & 27.88 & 0.846 & 0.207 & 5.13 \\
\hline
\end{tabular}
\caption{Per-scene result for ours (4DGS) on HyperNeRF Dataset}
\label{table:hypernerf1}
\end{table}
\begin{table}[h]
\centering
\begin{tabular}{llcccc}
\hline
Rate       & Name               & PSNR  & SSIM & LPIPS  & Size \\
\hline
\multirow{8}{*}{High} 
&bouncingballs & 41.00 & 0.994 & 0.016  & 2.03 \\
&hellwarrior   & 28.80 & 0.971 & 0.041  & 1.83 \\
&hook          & 32.52 & 0.974 & 0.030  & 1.99 \\
&jumpingjacks  & 35.37 & 0.984 & 0.021  & 1.70 \\
&lego          & 25.15 & 0.935 & 0.060  & 2.08 \\
&mutant        & 37.51 & 0.987 & 0.018 & 1.93 \\
&standup       & 38.25 & 0.99  & 0.015 & 1.67 \\
&trex          & 34.39 & 0.985 & 0.023 & 2.09 \\
\hline
\multirow{8}{*}{Mid} 
&bouncingballs & 40.91 & 0.994 & 0.016 & 1.95 \\
&hellwarrior   & 28.75 & 0.971 & 0.041 & 1.71 \\
&hook          & 32.59 & 0.974 & 0.031 & 1.76 \\
&jumpingjacks  & 35.02 & 0.985 & 0.022 & 1.62 \\
&lego          & 25.17 & 0.935 & 0.061 & 1.97 \\
&mutant        & 37.25 & 0.986 & 0.019 & 1.86 \\
&standup       & 37.95 & 0.990 & 0.015 & 1.59 \\
&trex          & 34.24 & 0.985 & 0.023 & 1.98 \\
\hline
\multirow{8}{*}{Low} 
&bouncingballs & 40.84 & 0.994 & 0.017 & 1.72 \\
&hellwarrior   & 28.9  & 0.971 & 0.042 & 1.67 \\
&hook          & 32.49 & 0.973 & 0.032 & 1.64 \\
&jumpingjacks  & 34.98 & 0.984 & 0.022 & 1.57 \\
&lego          & 25.13 & 0.934 & 0.063 & 1.81 \\
&mutant        & 37.08 & 0.986 & 0.020 & 1.73 \\
&standup       & 37.89 & 0.989 & 0.016 & 1.55 \\
&trex          & 34.17 & 0.984 & 0.024 & 1.83 \\
\hline
\end{tabular}
\caption{Per-scene result for ours (4DGS) on D-NeRF Dataset}
\label{table:dnerf1}
\end{table}

\begin{table}[h]
\centering
\begin{tabular}{llcccc}
\hline
Rate       & Name               & PSNR  & LPIPS  & DSSIM  & Size \\
\hline
\multirow{6}{*}{High} 
           & cut\_roasted\_beef & 32.95 & 0.058  & 0.0135 & 3.7  \\
           & cook\_spinach      & 32.37 & 0.054  & 0.0138 & 3.4  \\
           & coffee\_martini    & 28.84 & 0.089  & 0.0223 & 2.9  \\
           & flame\_salmon\_1   & 29.29 & 0.086  & 0.0208 & 3.3  \\
           & flame\_steak       & 33.04 & 0.045  & 0.0112 & 3.0  \\
           & sear\_steak        & 32.72 & 0.044  & 0.0107 & 2.8  \\
\hline
\multirow{6}{*}{Mid} 
           & cut\_roasted\_beef & 32.87 & 0.057  & 0.0133 & 2.8  \\
           & cook\_spinach      & 32.26 & 0.059  & 0.0149 & 2.8  \\
           & coffee\_martini    & 28.83 & 0.088  & 0.0212 & 2.5  \\
           & flame\_salmon\_1   & 29.17 & 0.089  & 0.0214 & 2.6  \\
           & flame\_steak       & 32.67 & 0.047  & 0.0114 & 2.2  \\
           & sear\_steak        & 32.25 & 0.045  & 0.0110 & 2.3  \\
\hline
\multirow{6}{*}{Low} 
           & cut\_roasted\_beef & 32.60  & 0.057  & 0.0137 & 2.4  \\
           & cook\_spinach      & 32.16 & 0.058  & 0.0146 & 2.4  \\
           & coffee\_martini    & 28.78 & 0.092  & 0.0221 & 2.4  \\
           & flame\_salmon\_1   & 28.97 & 0.089  & 0.0219 & 2.5  \\
           & flame\_steak       & 31.76 & 0.047  & 0.0118 & 1.9  \\
           & sear\_steak        & 31.57 & 0.048  & 0.0120 & 2.1  \\
\hline
\end{tabular}
\caption{Per-scene result for ours (4DGS) on Neu3D Dataset}
\label{table:dynerf1}
\end{table}

\begin{table}[h]
\centering
\begin{tabular}{llcccc}
\hline
Rate       & Name               & PSNR  & SSIM  & LPIPS  & Size \\
\hline
\multirow{8}{*}{High} 
&bouncingballs & 36.07 & 0.990  & 0.018 & 3.9  \\
&hellwarrior   & 37.10 & 0.969 & 0.048 & 2.8  \\
&hook          & 36.00 & 0.981  & 0.021 & 6.5  \\
&jumpingjacks  & 33.65 & 0.979 & 0.023 & 3.8  \\
&lego          & 24.89 & 0.934 & 0.054 & 6.5  \\
&mutant        & 39.56 & 0.990  & 0.011 & 4.7  \\
&standup       & 41.82 & 0.992 & 0.010 & 3.7  \\
&trex          & 30.16 & 0.977 & 0.025 & 12.9 \\
\hline
\multirow{8}{*}{Mid} 
&bouncingballs & 35.63 & 0.990 & 0.019 & 3.3  \\
&hellwarrior   & 36.98 & 0.970 & 0.050 & 1.9  \\
&hook          & 35.63 & 0.979 & 0.022 & 5.8  \\
&jumpingjacks  & 33.42 & 0.979 & 0.025 & 3.1  \\
&lego          & 24.88 & 0.935 & 0.055 & 5.6  \\
&mutant        & 39.54 & 0.990 & 0.011 & 4.0  \\
&standup       & 42    & 0.992 & 0.010 & 3.0  \\
&trex          & 30.19 & 0.976 & 0.025 & 12.2 \\
\hline
\multirow{8}{*}{Low} 
&bouncingballs & 35.39 & 0.990 & 0.020 & 3.0  \\
&hellwarrior   & 36.95 & 0.970 & 0.054 & 1.5  \\
&hook          & 35.73 & 0.979 & 0.022 & 5.4  \\
&jumpingjacks  & 33.49 & 0.979 & 0.025 & 2.7  \\
&lego          & 24.96 & 0.936 & 0.054 & 5.4  \\
&mutant        & 39.33 & 0.990 & 0.012 & 3.8  \\
&standup       & 41.65 & 0.992 & 0.01  & 2.7  \\
&trex          & 30.03 & 0.972 & 0.027 & 10.8 \\
\hline
\end{tabular}
\caption{Per-scene result for ours (SaroGS) on D-NeRF Dataset}
\label{table:dnerf2}
\end{table}

\begin{table}[h]
\centering
\begin{tabular}{llcccc}
\hline
Rate       & Name               & PSNR  & LPIPS  & DSSIM  & Size \\
\hline
\multirow{6}{*}{High} 
&cut\_roasted\_beef             & 33.59 & 0.039 & 0.0113 & 18.6 \\
&cook\_spinach          & 32.92 & 0.042 & 0.0130  & 21.7 \\
&coffee\_martini           & 28.94 & 0.060 & 0.0211 & 27.6 \\
&flame\_salmon\_1 & 28.99 & 0.061 & 0.0217 & 18.7 \\
&flame\_steak     & 33.56 & 0.034 & 0.0109 & 26.7 \\
&sear\_steak      & 33.57 & 0.034 & 0.0103 & 12.4 \\
\hline
\multirow{6}{*}{Mid} 
           & cut\_roasted\_beef & 33.46 & 0.043 & 0.0122 & 17.6 \\
           & cook\_spinach      &32.80  & 0.042 & 0.0132  & 19.9 \\
           & coffee\_martini    & 28.61 & 0.066 & 0.0226 & 23.8 \\
           & flame\_salmon\_1   & 28.77 & 0.064 & 0.0217 & 15.8 \\
           & flame\_steak      & 33.29 & 0.035 & 0.0113 & 23.7 \\
           & sear\_steak        & 33.47 & 0.039 & 0.0114 & 10.3 \\
\hline
\multirow{6}{*}{Low} 
           & cut\_roasted\_beef & 33.1  & 0.042  & 0.0122 & 16.0  \\
           & cook\_spinach      & 32.78 & 0.042  & 0.0131 & 17.7  \\
           & coffee\_martini    & 28.62 & 0.064  & 0.0224 & 22.0  \\
           & flame\_salmon\_1   & 28.58 & 0.064  & 0.0218 & 15.2  \\
           & flame\_steak       & 33.18 & 0.036  & 0.0119 & 22.4  \\
           & sear\_steak        & 33.31 & 0.048  & 0.0113 & 8.9  \\
\hline
\end{tabular}
\caption{Per-scene result for ours (SaroGS) on Neu3D Dataset}
\label{table:dynerf2}
\end{table}

\clearpage

\label{sec:rd}


\end{document}